# Conformer diversity over quantity in COSMO-RS-based LLE predictions

*Ilah Joos[a], Simon Müller[b], Jeriffa De Clercq[a], Irina Smirnova[b], Maarten K. Sabbe[a,c,*], Jeroen Lauwaert[a]*

[a] Industrial Catalysis and Adsorption Technology, Ghent University, 9000 Ghent, Belgium.

[b] Institute of Thermal Separation Processes, Hamburg University of Technology, 21073 Hamburg, Germany.

[c] Laboratory for Chemical Technology, Ghent University, 9052 Ghent, Belgium.

*Corresponding author: Maarten.Sabbe@UGent.be

**Abstract**

Accurate phase equilibrium prediction is vital for chemical process design, particularly for challenging systems involving complex intermolecular interactions. This study examines how conformer selection and sigma profiles affect COSMO-RS predictions of liquid–liquid equilibria (LLE's), using the tetrahydrofurfuryl alcohol (THFA)–methyl isobutyl ketone (MIBK)–water system as a case study. THFA's conformational flexibility and the contrasting polarity of the two phases, make this system highly sensitive to conformer effects.

Simulations were performed at two levels of theory: TZVP and TZVPD-FINE. The TZVP calculations failed to reproduce the experimentally observed LLE behavior, whereas the TZVPD-FINE calculations showed strong agreement. This contrast arises because TZVP favors the lowest-energy, less polar conformers, while TZVPD-FINE emphasizes more polar, hydrogen-bonding-capable conformers. Single-conformer simulations showed that individual conformers can skew predictions due to inaccurate sigma profile representation and that only a representative conformer set yields reliable predictions. Comparative simulations further indicated that predictive accuracy depends more on sigma-profile diversity than the number of conformers used.

These findings highlight the dual importance of conformer representation and the application of accurate sigma profiles in COSMO-RS. For systems governed by subtle interactions such as liquid-liquid equilibria, both aspects must be carefully considered to ensure predictive reliability. These insights increase understanding of COSMO-RS and help towards its correct application in both academic and industrial contexts.

## Introduction

Accurate prediction of phase equilibria is essential in chemical process design, separation technology, and solvent screening. To this end, a wide range of predictive activity coefficient models have been developed.[1–8] Among these, Conductor-like Screening Model for Real Solvents (COSMO-RS) has emerged as a widely adopted, efficient, and often reliable approach for predicting phase equilibria and thermodynamic properties in general.[9–20] Thanks to its performance, it is increasingly applied as a complement to, or even a substitute for, experimental data. Recent studies illustrate its versatility: Silva et al. successfully predicted activity coefficients for polymer solutions;[16] Chagnoleau et al. applied COSMO-RS for the accurate predictions of binodal curves and partition coefficients of binary systems;[17] Wang et al. evaluated its potential for prediction liquid-liquid equilibria (LLE) containing deep eutectic solvents;[18] and in our previous work, we demonstrated that COSMO-RS can effectively replace vapour-liquid equilibrium (VLE) data for furfural derivatives.[19]

Despite these successes, COSMO-RS simulations are not always correct. In particular, the quality of COSMO-RS predictions depends largely on how accurately the molecular interactions are represented. These interactions are captured in sigma surfaces derived from quantum chemical calculations, which encode the electrostatic and hydrogen-bonding properties of each molecule. A key aspect in generating accurate sigma surfaces is the selection of molecular conformers, since each conformer contributes a distinct sigma surface to the overall model.[16,21,22] Numerous studies have emphasized the importance of constructing a representative and accurate conformer set to ensure reliable predictions. These sets must include all relevant conformers for each phase under consideration, to capture the full thermodynamic behavior across different phases.[16,19,21,23–25]

The two liquid phases in LLE's are often very different in nature, making these systems especially challenging to predict accurately.[9,13,18] Because LLE's are highly sensitive to subtle variations in intermolecular interactions, the reliability of predictions depends strongly on the quality and representativeness of the conformers used. Recent studies have shown that not only the selection of conformers but also how they are weighted within the COSMO-RS framework can strongly affect the predicted phase behaviour.[9,16,25] For example, Cordova et al. demonstrated that adjusting the conformer weighting can lead to substantially different phase equilibrium predictions. By tailoring these weightings to better reflect experimental observations, they achieved more accurate phase equilibrium predictions.[9] This highlights the critical role of conformer weighting in enhancing the reliability of COSMO-RS simulations for LLE systems and phase equilibria in general.

To gain a deeper understanding of how conformer weighting and sigma profiles steer phase equilibrium predictions, this study presents an in-depth analysis of COSMO-RS simulations for the ternary tetrahydrofurfuryl alcohol (THFA)-methyl isobutylketon (MIBK)-water system. This system was selected because the conformational flexibility of THFA, particularly its rotable alcohol group, can alter its hydrogen-bonding capacity and, consequently, its interactions with both MIBK and water. The focus on LLE is intentional, as these phase equilibria are especially sensitive to conformer effects.[9,13,18] Accordingly, these simulations can provide valuable insights into COSMO-RS.

The impact of the sigma profiles and conformer weighting is assessed by comparing the results from Triple-Zeta Valence with Polarization (TZVP) and Triple-Zeta Valence with Polarization and Diffuse functions-FINE (TZVPD-FINE) simulations in COSMOthermX, offering additional perspectives on the strengths and limitations of different COSMO-RS parameterizations.

Ultimately, the insights presented here aim to enhance the understanding of COSMO-RS simulations and their reliability, particularly in the context of liquid–liquid equilibria. This contributes to more informed and confident use of the model in both academic and industrial settings.

## Materials and methods

### *Chemicals*

Tetrahydrofurfuryl alcohol (THFA) was supplied by Fischer Scientific, methyl isobutyl ketone (MIBK) by Merck Life Science and ethanol by Chemlab Analytical. Demineralized water was used throughout this work. All chemicals had a purity of at least 99 wt% and were used without further purification.

### *Liquid-liquid experiments*

To determine the tie-lines of the THFA-MIBK-Water mixture, several mixing points within the miscibility gap were prepared in separation funnels. The mixtures were vigorously shaken and then allowed to equilibrate overnight at 298.15 K. The composition of the resulting organic and aqueous samples were analysed using gas chromatography (GC) and Karl Fischer titration to determine the THFA/MIBK ratio and the water content respectively.

For the GC analysis a Shimadzu GC2010-Pro equiped with a flame ionization detector (FID) was employed. Nitrogen served as the carrier gas, and the stationary phase was a SH-Stabilwax column (30 m × 0.32 mm × 0.50 μm). The following temperature program was applied: the GC oventemperature was first held at 313.15 K for 1 min, then increased at a rate of 20 K/min to 403.15 K, where it was maintaned for another 1 min. The Karl Fischer analysis was performed using a coulometric MKC 501 apparatus, with Hydranal Coulomat AK and Coulomat CG-K as the anolyte and catholyte, respectively.

### *COSMOthermX*

Conformer sets were generated using COSMOconf v3.0. Calculations were performeed with either the TZVP or TZVPD parmeterizations were performed. In COSMOthermX, TZVPD are always paired with the FINE option, which provides a refined cavity construction.[26] In general, TZVP calculations are faster, whereas TZVPD-FINE calculations offer higher accuracy. Using the σ-profiles from COSMOconf LLE simulations were carried out using the BIOVA COSMOtherm 2020 software packag. All simulations were performed for isothermic phase equilibria at 298.15 K and always applying the same conformers for both MIBK and water. The midpoint of every experimental tie-line was used as the input mixing point for the simulations. For each liquid phase, the mole fractions as well as the conformer weights were obtained for all three components.

### *Comparison of the LLE simulations with the experimental data*

The agreement of the simulated LLE's with the experimentally observed ones is assessed by means of their mean percentage deviation (MPD), for tie-lines with the same mixing point, for both the organic and the aqueous phases (equation 1 and 2 respectively). The model simulations are considered to be in good agreement with the experimental data when the MPD is less than or equal to 5%.

$$MPD_{org} = \frac{\sum_{i=1}^{n}\left(\left|w'^{exp}_{component,i} - w'^{sim}_{component,i}\right| + \left|w'^{exp}_{organic\,solvent,i} - w'^{sim}_{organic\,solvent,i}\right| + \left|w'^{exp}_{water,i} - w'^{sim}_{water,i}\right|\right)}{n\sqrt{2}} 100 \leq 5 \qquad (1)$$

$$MPD_{aq} = \frac{\sum_{i=1}^{n}\left(\left|w''^{exp}_{component,i} - w''^{sim}_{component,i}\right| + \left|w''^{exp}_{organic\,solvent,i} - w''^{sim}_{organic\,solvent,i}\right| + \left|w''^{exp}_{water,i} - w''^{sim}_{water,i}\right|\right)}{n\sqrt{2}} 100 \leq 5 \qquad (2)$$

In equation 1, for the experimental and simulated tie-lines, $w'_{component,i}$, $w'_{organic\ solvent,i}$ and $w'_{water,i}$ represent the mass fractions of the component, organic solvent and water respectively in the organic phase. Analogously, in equation 2, $w''_{component,i}$ , $w''_{organic\ solvent,i}$ and $w''_{water,i}$ represent the corresponding mass fractions in the aqueous phase. The number of tie-lines is givenn by n. Since mass fractions are values between 0 to 1, the maximum distance between two compositions in these ternary systems is √2.

**Results and discussion**

The predictive performance of COSMO-RS is evaluated using conformers generated by COSMOconf at either the TZVP or TZVPD-FINE levels of theory. For both methods, the same THFA conformer set is applied (Figure 1) with the energy of the conformer in an ideal conductor increasing from conformer c000 to conformer c007. As shown in Figure 2a and Table 1, the simulation using the TZVP method gives a miscibility gap, but failed to reproduce the experimental liquid–liquid partitioning behavior. The solubilities are off and the predicted orientation of the tie lines was nearly opposite to that observed experimentally, indicating a mismatch in the predicted phase behavior. In contrast, when the same conformer set was used in TZVPD-FINE simulations (Figure 2b, Table 1), the predicted LLE diagrams agreed much better with the experimental data, both the solubilities and the orientation of the tie lines were better described. The MPD of the aqueous phase is smaller than 5%, confirming that the experimental LLE is well predicted. However, this is not true for the organic phase, which with an MPD of 5.27% slightly surpasses the treshold.

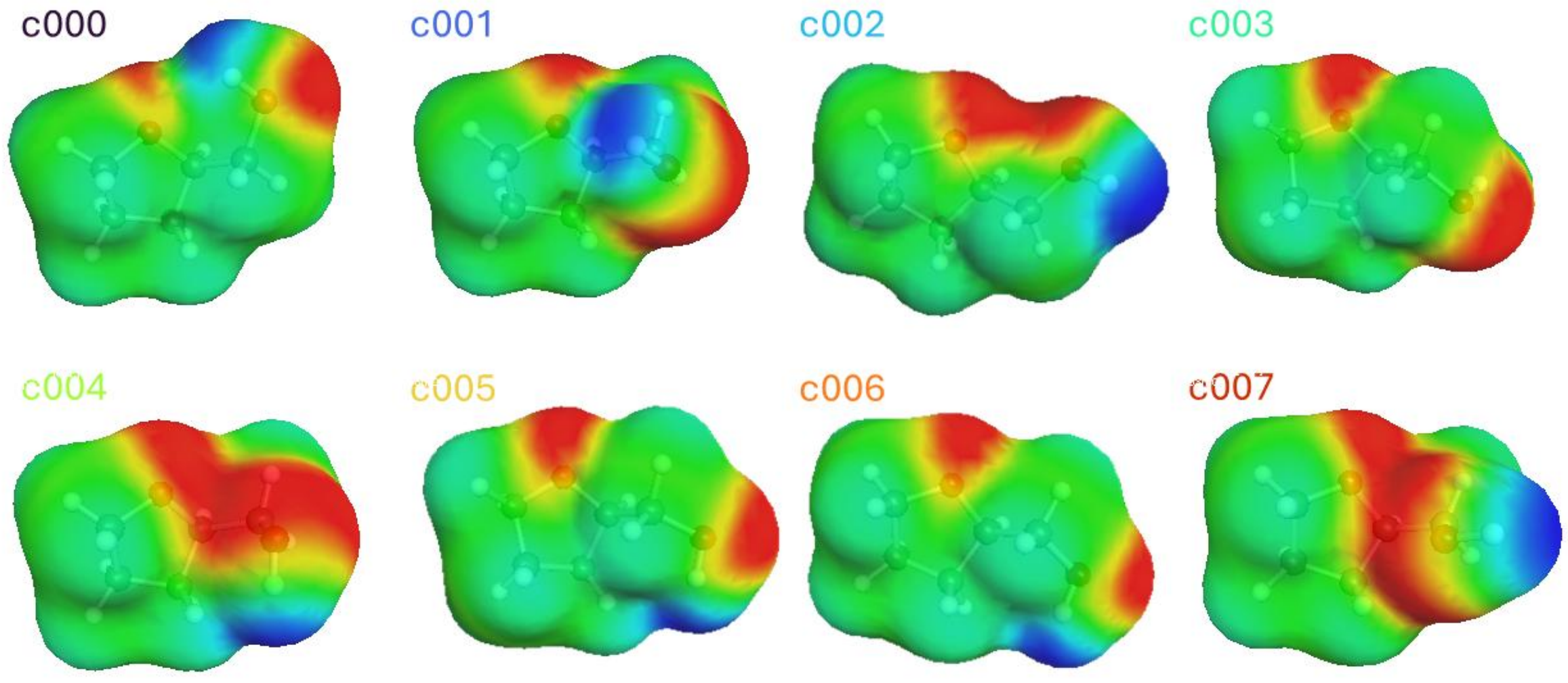

**Figure 1**. TZVPD-FINE conformers of THFA used in COSMOthermX: c000 (dark violet), c001 (dark blue), c002 (light blue), c003 (green), c004 (lime green), c005 (yellow), c006 (orange) and c007 (red). The conformer set was generated with COSMOconf v3.0.

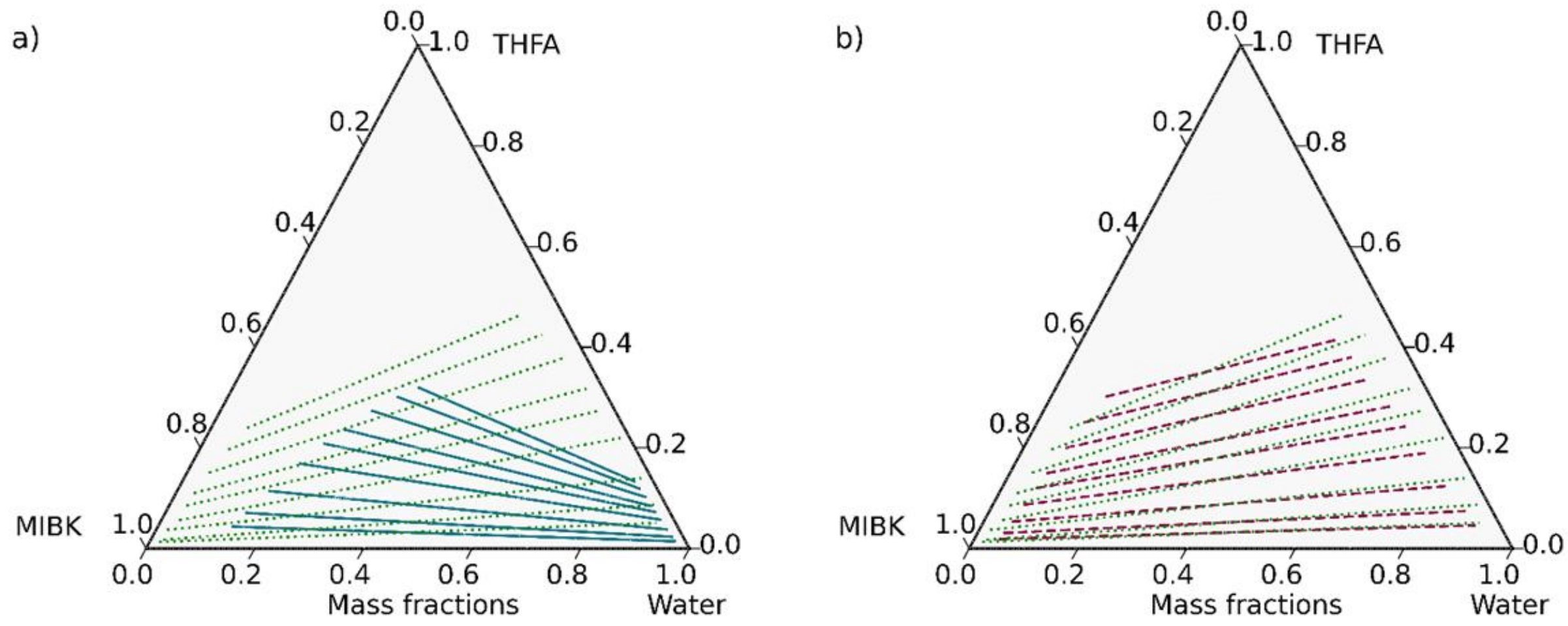

**Figure 2.** Experimental LLE data (dotted green, Table S25) and a) the TZVP (solid teal, Table S15) simulation in COSMOthermX, and b) the TZVPD-FINE (dashed red, Table S1) simulation in COSMOthermX of the THFA/MIBK/water mixture.

**Table 1.** Mean percentage deviations for the organic (MPDorg) and aqueous (MPDaq) phase, between the simulated and experimental LLE's.

| | THFA conformers applied | $MPD_{org}$ (%) | $MPD_{aq}$ (%) |
|---|---|---|---|
| TZVP | Full conformer set | 23.81 | 17.61 |
| | c000 | 22.90 | 20.44 |
| | c002 | 18.30 | 3.29 |
| TZVPD-FINE | Full conformer set | 5.27 | 3.56 |
| | c000 | 9.00 | 5.67 |
| | c002 | 3.56 | 5.21 |
| | c000 & c002 | 3.89 | 4.10 |

To understand the origin of the distinct difference between the TZVP and TZVPD-FINE predictions, the conformer weights of THFA in MIBK and water, at infinite dilution (Figure 3, Table S2 and S16), as well as the corresponding sigma profiles (Figure 4) were compared for both simulations. This analysis revealed that different conformers dominate in the two methods. In the TZVP calculations, the most abundant conformer was the lowest-energy structure (c000, Figure 1) which, compared with most other conformers, exhibits a relatively non-polar sigma profile (Figure 4). This aligns with the observations that in the TZVP prediction THFA preferentially partitions into MIBK rather than water. Conversely, the TZVPD-FINE prediction favored more polar conformers. TZPVD-FINE seems suited for capturing interactions such as van der Waals forces and hydrogen bonding. Consequently, it does not preferentially select the lowest energy conformer, but rather those that are more polar and feature accessible functional groups capable of hydrogen bonding. This effect is particularly pronounced in the aqueous phase, where the more polar conformer c002 (Figure 1) accounts for nearly 50% of the conformer distribution.

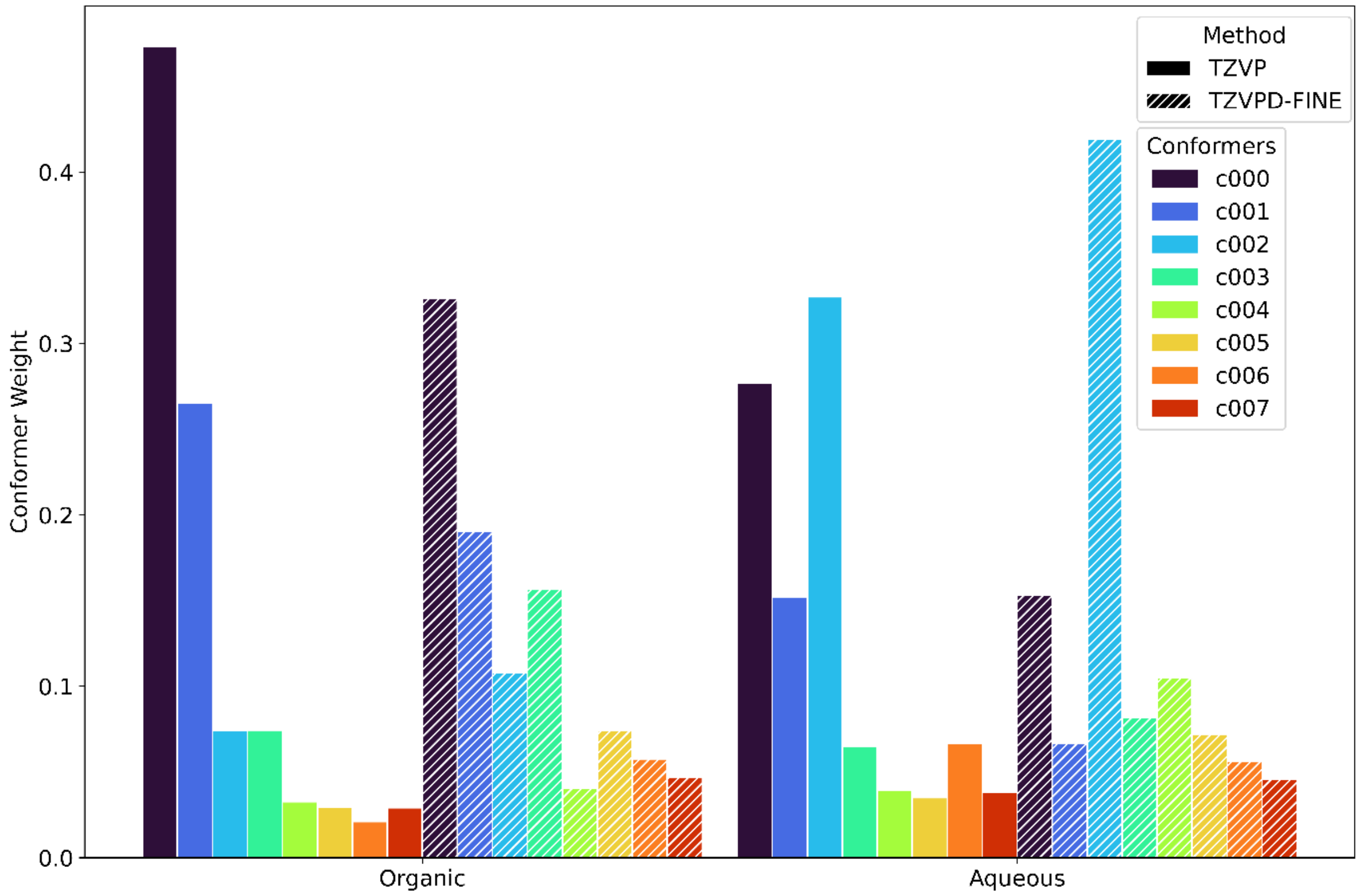

**Figure 3.** THFA conformer distribution in the organic (MIBK-rich) and aqueous (water-rich) phase for the TZVP and TZVPD-FINE simulations in COSMOthermX.

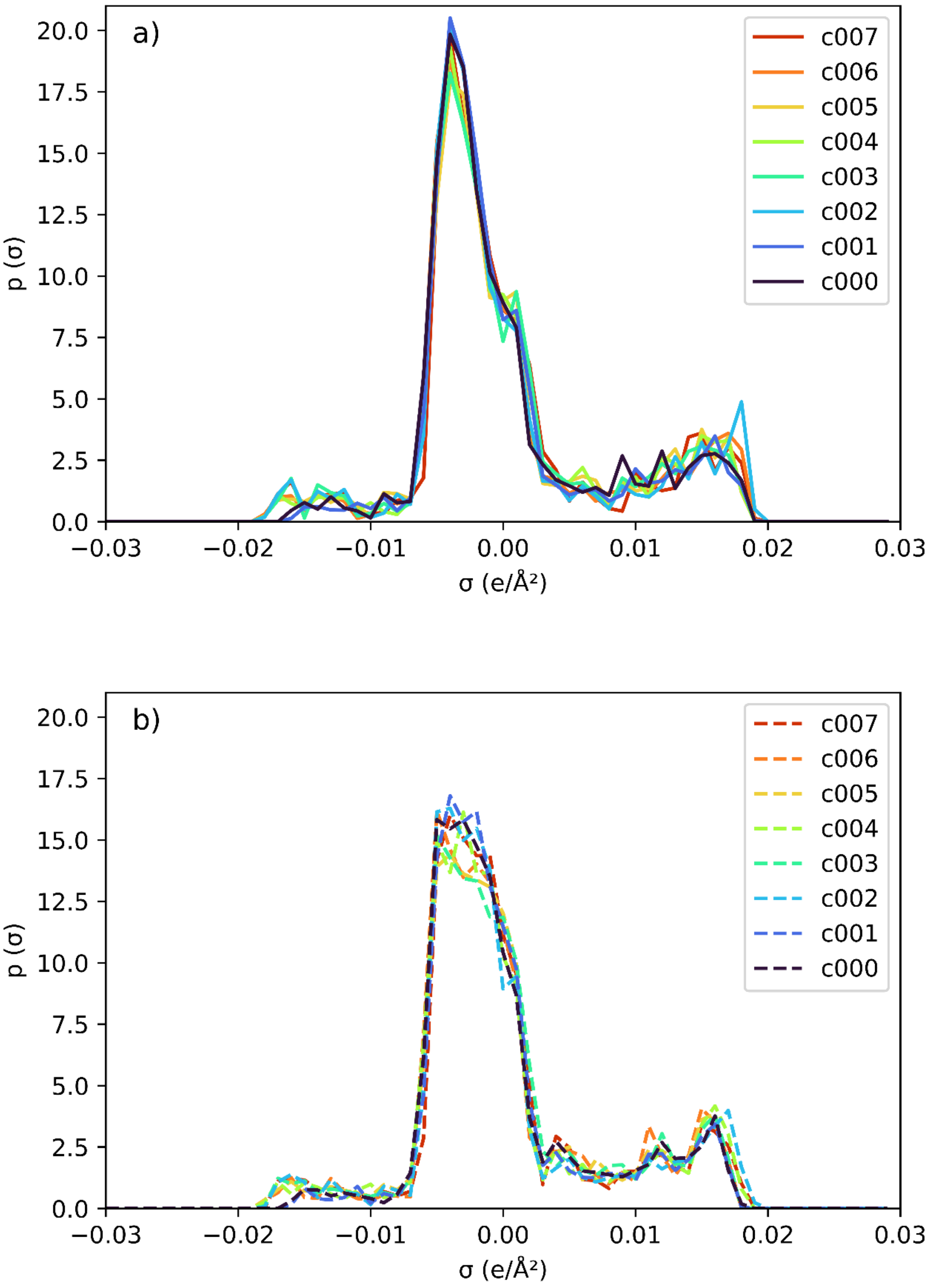

**Figure 4.** Sigma profiles of the TZVP (a) and TZVPD-FINE (b) conformers of THFA.

To isolate the effect of conformer choice, single-conformer simulations were performed, using the dominant conformer from each method, i.e., c000 and c002. Figure 5 compares the resulting LLE predictions and the MPD's are given in Table 1, while additional single conformer simulations, the full LLE data and the respective MPD's are provided in the Supporting Information (Figures S1-S2, Tables S3-S10, Tables S17-S24 and Table S26). The results reveal that for both TZVP and TZVPD-FINE, the orientation of the tie-lines in the LLE diagrams depends on the conformer used. Simulations with c000, the dominant TZVP conformer, produce tie-line orientations more similar to the original TZVP results, calculated with the full set of conformers, with THFA preferentially partitioning into the organic phase. In contrast simulations using c002, the main TZVPD-FINE conformer, show tie-lines favouring the aqueous phase, with slopes shifting the slopes in the opposite direction, positively affecting the MPD.

When all conformers from the set are included, their opposing effects are balanced out, as seen in the multi-conformer simulations shown in Figure 2. Since the orientation of the tie-lines is worse than for the full conformerset simulation, for the c000-based simulation both MPD's surpass the treshold. However, the c002-based simulation shifts the tie-lines in the opposite direction, with a similar deviation as the simulation applying the full conformer set. This is confirmed by the MPD's which have similar values. Since it is hard to predict which conformer might give good results, even with the more accurate TZVPD-FINE method, relying on a single, and thus less representative conformer is not advisable. Selecting the wrong conformer may lead to a poor predictive performance.

Further comparison of the results for each conformer across both methods also revealed that even when the same conformer is used, the predicted phase behaviour differs substantially between TZVP and TZVPD-FINE. Notably, using TZVP, even the most polar conformers fail to reproduce the correct tie-line slopes (Figure 5, Table 1), with THFA favouring the aqueous phase, highlighting the method's limitation. This points to a fundamental difference between the two approaches when applied to the system in question: TZVPD-FINE places greater emphasis on conformers with enhanced hydrogen-bonding capabilities, whereas TZVP tends to prioritise the lowest-energy structures, irrespective of their interaction potential.

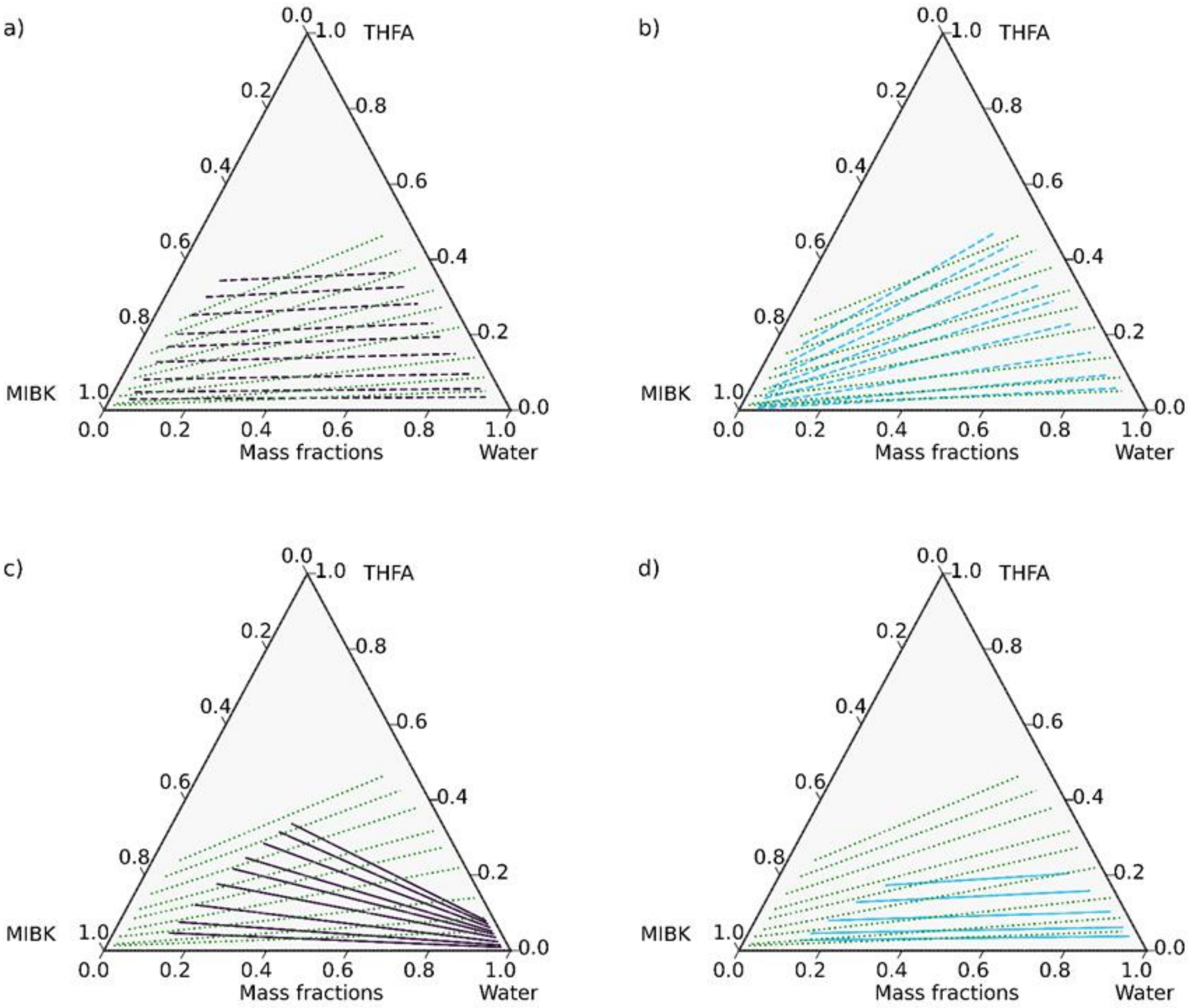

**Figure 5**. Comparison of the experimental LLE data (dotted green, Table S25) with the TZVPD-FINE (dashed) simulations with single conformers of THFA: a) c000 (dark violet, Table S3), b) c002 (light blue, Table S5) and the TZVP (solid) simulations with single conformers of THFA: c) c000 (dark violet, Table S17), d) c002 (light blue, Table S19) for the THFA/MIBK/water mixture.

For complex systems such as the THFA–MIBK–water mixture, these findings demonstrate that accurate phase equilibrium predictions require not only a representative conformer set but also a computational method capable of capturing subtle intermolecular interactions. Simply adjusting conformer weighting is insufficient if the underlying method lacks the resolution to distinguish interaction strengths. Consequently, for the system studied TZVPD-FINE proofs to be essential for obtaining reliable COSMO-RS predictions in such systems where the hydrogen bonding interaction decides whether or not a conformer is dominant.

Comparison of the sigma profiles of the different THFA conformers (Figure 4) reveals two distinct groups. Conformers c000 and c001 have similar, relatively non-polar sigma profiles, while conformers c002-c007 form a second group with similar sigma profiles, characterized by a higher polarity. The higher polarity in this latter group arises from better accessibility of the alcohol group for hydrogen bonding, which directly impacts the LLE predictions.

To explore this further, a two-conformer simulation was performed at the TZVPD-FINE level, using only the lowest-energy conformer from each group (c000 and c002, Figure 6). Surprisingly, for the system studied, this simplified model yielded more accurate LLE predictions than the full conformer set generated by COSMOconf. Unlike for the prediction with the full conformer set, for the c000-c002 prediction both MPD's are smaller than 5% (Table 1), confirming that this simulation predicts the experimentally observed LLE well. This suggests that capturing the diversity of relevant sigma profile types might be more critical than including many conformers. A carefully selected minimal set of representative conformers may thus provide better predictive performance than an extensive but less targeted ensemble.

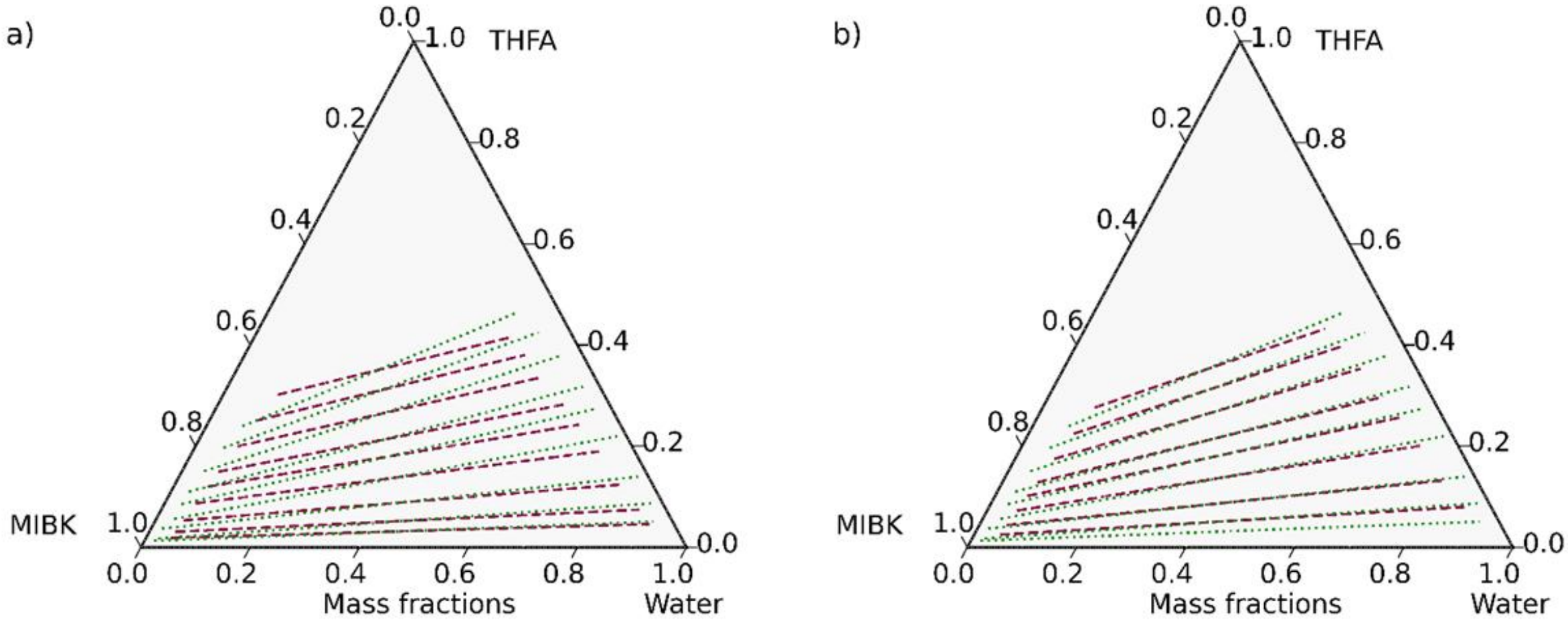

**Figure 6**. Comparison of the experimental LLE data (dotted green, Table S25) and the TZVPD-FINE (dashed red) simulations with a) the full conformer set (Table S1) and b) conformers c000 and c002 (Table S11) of THFA.

As a test of the generality of these observations, a similar two-conformer simulation was performed for a second ternary system: 1,4-butanediol, dimethyl succinate, and water.[27] Like the THFA–MIBK–water system, this mixture involves components with hydrogen bonding potential and polarity contrasts arising conformational differences. The sigma surfaces and sigma profiles are provided in the Supporting information (Figure S3-S4). Using a simplified model with just the lowest-energy conformer from each sigma profile type, i.e. one more and one less polar, for 1,4-butanediol, again produced LLE predictions that are comparable to, but slightly better than those with the entire conformer set (Figure 7, Table 2), the conformer weights of the simulation with the entire conformer set are given in Table S13**.** Notably, due to a poor prediction of the solulbility, the $MPD_{aq}$ still surpasses the treshold. Nonetheless, the consistency of these results reinforce the hypothesis that accurately capturing the diversity of sigma profile types outweighs simply increasing the number of conformers. Nonetheless, while promising, these findings should be verified across a broader range of systems to confirm their general applicability. In particular, further studies are needed to determine whether this approach holds for systems with different molecular features or interaction types.

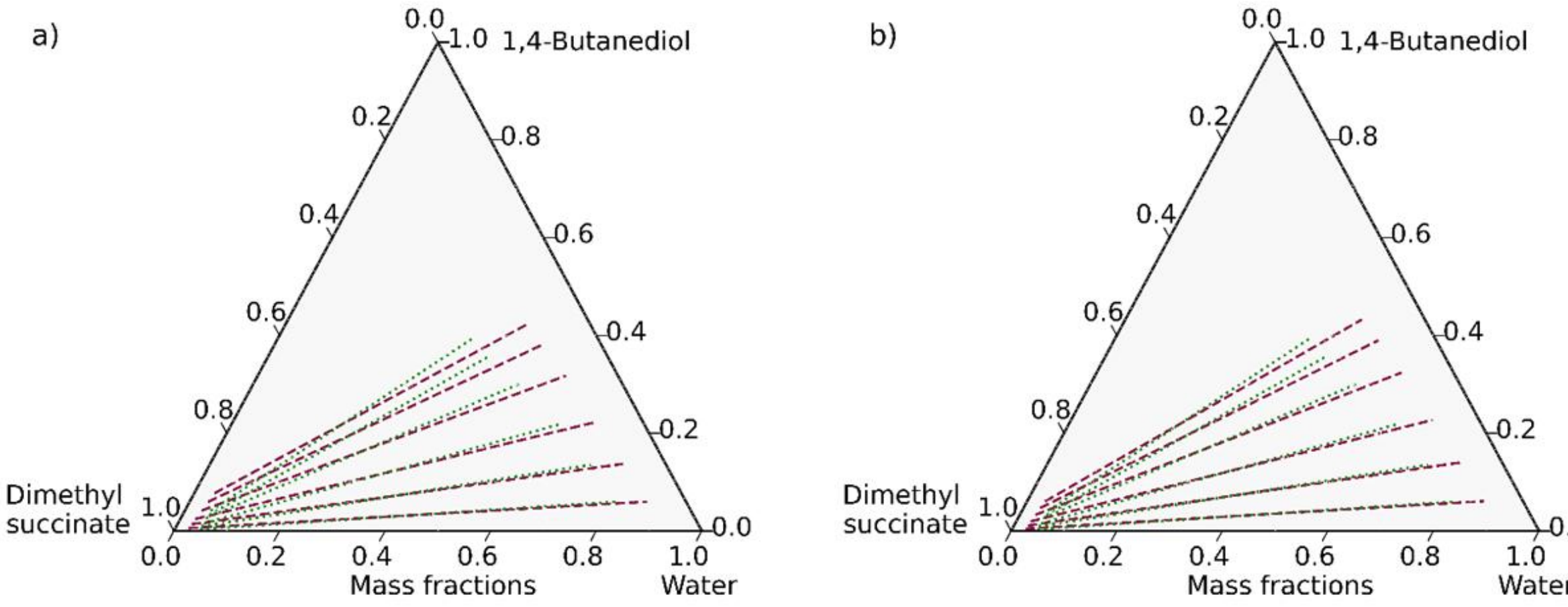

**Figure 7.** Comparison of the experimental 1,4-butanediol-dimethyl succinate-water LLE data[27] (dotted green) with the TZVPD-FINE (dashed red) simulations with a) the full conformer set (Table S12) and b) conformers c000 and c002 (Table S14) of 1,4-butanediol.

**Table 2.** Mean percentage deviations (MPD's) for the organic and aqueous phase, between the simulated and experimental LLE's for the 1,4-butanediol-dimethyl succinate-water mixture.

| | 1,4-butanediol conformers applied | $MPD_{org}$ (%) | $MPD_{aq}$ (%) |
|---|---|---|---|
| TZVPD-FINE | Full conformer set | 2.17 | 7.95 |
| | c000 & c002 | 1.90 | 7.73 |

## Conclusion

This study sets out to deepen the understanding of how conformer weighting and sigma profiles affect phase equilibrium predictions within the COSMO-RS framework, using the ternary THFA-MIBK-water system as a case study. The conformational flexibility of THFA, particularly the rotable alcohol group, makes it a sensitive component for evaluating hydrogen bonding interactions and their impact on LLE predictions. By comparing simulations performed with TZVP and TZVPD-FINE levels of theory, the impact of both conformer selection and weighting on the reliability of COSMO-RS predictions has been assessed.

It has been demonstrated that reliable LLE predictions require not only selecting representative conformers, but also careful attention to conformer weighting and the quality of the sigma profiles, whicht depends on the level of the computational method applied. Neglecting either, i.e., by relying on a less representative conformer or a lower-level computational method, leads to poor agreement with experimental data. Single-conformer simulations highlighted how individual conformers can skew predictions, and how their combined effects counterbalance each other. Similar results were obtained for a second ternary system (1,4-butanediol-dimethyl succinate-water) suggesting broader relevance. However, further validation across diverse systems is essential to confirm the generality of these findings.

Ultimately, this work underscores that the predictive reliability of COSMO-RS for complex systems hinges on the interplay between conformer diversity and methodological accuracy, particularly for systems governed by complex and sensitive intermolecular interactions such as LLE. Recognizing and addressing the dual importance of sigma profile accuracy and selecting accurate conformer sets will enable more robust application of COSMO-RS in both academic research and industrial process design.

**Acknowledgements**

This work was supported by the Research Foundation-Flanders (FWO) through Travel Grant V402625N. The authors would like to thank Julia Verlodt for her help with the experimental validation.